# Directly wireless communication of human minds via non-invasive brain-computer-metasurface platform

Qian Ma[1,3,4,†], Wei Gao[2,5,†], Qiang Xiao[1,3,4], Lingsong Ding[2,5], Tianyi Gao[2,5], Yajun Zhou[2,5], Xinxin Gao[1,3,4], Tao Yan[1,3,4], Che Liu[1,3,4], Ze Gu[1,3,4], Xianghong Kong[6], Qammer H. Abbasi[7], Lianlin Li[4,8], Cheng-Wei Qiu[6*], Yuanqing Li[2,5*] and Tie Jun Cui[1,3,4*]

[1] Institute of Electromagnetic Space, Southeast University, Nanjing 210096, China
[2] School of Automation Science and Engineering, South China University of Technology, Guangzhou 510641, China
[3] State Key Laboratory of Millimeter Wave, Southeast University, Nanjing 210096, China
[4] Center of Intelligent Metamaterials Pazhou Laboratory, Guangzhou 510330, China
[5] Research Center for Brain-Computer Interface, Pazhou Lab, Guangzhou 510330, China
[6] Department of Electrical and Computer Engineering, National University of Singapore, Singapore
[7] University of Glasgow, James Watt School of Engineering, Glasgow, G12 8QQ, United Kingdom
[8] State Key Laboratory of Advanced Optical Communication Systems and Networks, Department of Electronics, Peking University, 100871 Beijing, China

[†]Qian Ma and Wei Gao: These authors contributed equally to this work.
*Co-corresponding authors: tjcui@seu.edu.cn; auyqli@scut.edu.cn; chengwei.qiu@nus.edu.sg.

# Abstract

Brain-computer interfaces (BCIs), invasive or non-invasive, have projected unparalleled vision and promise for assisting patients in need to better their interaction with the surroundings. Inspired by the BCI-based rehabilitation technologies for nerve-system impairments and amputation, we propose an electromagnetic brain-computer-metasurface (EBCM) paradigm, regulated by human's cognition by brain signals directly and non-invasively. We experimentally show that our EBCM platform can translate human's mind from evoked potentials of P300-based electroencephalography to digital coding information in the electromagnetic domain non-invasively, which can be further processed and transported by an information metasurface in automated and wireless fashions. Directly wireless communications of the human minds are performed between two EBCM operators with accurate text transmissions. Moreover, several other proof-of-concept mind-control schemes are presented using the same EBCM platform, exhibiting flexibly-customized capabilities of information processing and synthesis like visual-beam scanning, wave modulations, and pattern encoding.

# Introduction

To directly inspect and distinguish human's will, brain-computer interface (BCI) is presented to establish the communication between brain and devices. By collecting spontaneous or specifically evoked electro-encephalography (EEG) signals from the scalp via non-invasive electrodes, BCI can decode operator's intentions and send commands to the controlled objects, without any requirements for the operator's muscle activity[1-3]. P300 potentials, steady-state visual evoked potentials[4] (SSVEPs), and sensorimotor rhythms are three typical brain patterns in EEG-based BCIs. The P300-based BCIs, which identify the operator's intention from the P300 potentials evoked by flashes of the corresponding buttons on the control panel, have been frequently used[5-7] to assist the handicapped[8] or medical applications[3].

Metamaterials and metasurfaces have showcased unparalleled electromagnetic (EM) regulating capabilities, enabling various novel phenomena on EM wave manipulations like the anomalous diffraction[9], invisibility cloaking[10,11], lensing[12], and imaging[13,14]. Recently, digital coding metasurfaces, incorporating with PIN diodes[15], varactors[16,17], micro-electromechanical systems[18] and amplifiers[19,20], have enabled active, real-time, and programmable controls over the EM functionalities [21-23], which used to be static or quite limited in conventional passive counterparts. This powers up abundant sub-directions like time-coding and space-time-coding metasurfaces[24,25], self-adaptively smart metasurfaces[26-28], programmable holograms[29,30], and direct information processing[31].

In this work, we cast a roadmap fusing the reprogrammable EM metasurfaces with BCIs. We firstly propose an electromagnetic brain-computer-metasurface (EBCM) to flexibly and non-invasively control the information syntheses and wireless transmissions. We then design and demonstrate the wireless text-communications between two BCI operators by using a 2-bit programmable metasurface, under the brain control. A P300-based BCI is applied to translate the operator's brain messages to EEG signals, and further to computer commands, which are fed into the electrical control system of the programmable metasurface. After that, the digital messages are wirelessly transmitted and received by two programmable metasurfaces to fulfill the text communication in experiments. In addition, three typical applications are also designed and experimentally verified, including visual-beam scanning, wave modulations, and pattern encoding. The measured data of multiple scattering modulations have shown good consistence with the operator's intention, which can be well reflected from the recorded EEG. We envision that the programmable metasurface system will become an incubator that can integrate the artificial intelligence and human brain intelligence to form more advanced intelligent systems.

## Results

**Principle of the EBCM**

We witnessed impressive advancements in the programmable metasurfaces [25,32,33]. On the other hand, BCI has been captivated a popular notion of enabling brain signals to directly deliver the physical actions. With non-invasive electrodes [5,6], the P300 potentials can be identified from the brain signals under a certain time sequence of visual stimuli. In this work, we intuitively combine the P300 coding characteristics with the programable information metasurface, and experimentally demonstrate a robust P300-based EBCM to showcase the robust control of EM wireless information (e.g., synthesis, manipulation, and encoding/decoding) with only inputs from the brain signals and the designed functions, as shown in Fig. 1a.

A displayer is placed in front of the operator to show the graphical user interface (GUI), and there is a virtual button matrix, as exhibited in Fig. 1b. Different buttons correspond to different coding-pattern operations of EBCM. Each trial corresponds to one command sending, and the buttons start to flash successively in a random order for about 5 rounds, each of which contains one flash for each button. The flash sequence is randomly generated before each trial. Fig. 1c presents the flashing sequence of 40 buttons (40 rows) in the beam-scanning scheme, where the yellow squares indicate the starting point of the stimulus. The vertical axis represents the number of buttons from 1 to 40, while the horizontal axis represents the 160 sequential stimulus flashes marked by yellow bars in Fig. 1c. Each marked block in Fig. 1c represents a duration of 30 ms and each flash lasts for 100 ms, which will span slightly more than 3 blocks. In Fig. 1c, we further show a zoom-in view of three flashes to illustrate the stimulus time sequences of different buttons more clearly.

The operator's attention is focused on the button corresponding to the command that he/she would like to issue (i.e., the target). When the target flashes, due to the oddball effect, a positive potential may be detected from EEG after approximately 300 ms, dubbed P300 potential[34]. Such a P300-based BCI has been experimentally proved effective for the brain to directly control the external devices [5,35]. To present the brain signals for two stimulus types (target and nontarget), we exhibit the measured EEG signals in Fig. 1d, where the red and blue curves correspond to the signals for target and nontarget stimulus, respectively. In each subfigure of Fig. 1d, the signals corresponding to both target and nontarget buttons are averaged across multiple flashes. For each button flash in a trial, a segment of EEG signal from 0 to 600 ms after the flash onset is extracted and corrected with a baseline extracted from 200 ms before the flash onset. Next, we exploit this sequential stimulus to directly control the field programmable gate array (FPGA) to execute the related instructions in the operators' mind, and successfully realize the paradigm shift from the brain signal to EM scattering manipulations. As a proof-of-concept demonstration of EBCM, we present wireless text communications between two operators by the minds. By designing specific rules between EEG and EM signals, the transmitted text in mind is accurately received and decoded in experiments. To exhibit more functions realized by the same EBCM, we also demonstrate three applications, which include 1) EM-beam scanning; 2) multiple scattering-beam switching; and 3) metasurface pattern encoding. More details of the EBCM platform are provided in Supplementary Note S1 [36].

**Wireless text communication by mind based on EBCM**

To fully exhibit the application of EBCM, we design and experimentally demonstrate wireless text communications by mind, as illustrated in Fig. 2a. A text GUI is provided for the BCI operator (see Supplementary Note S2 [36]), in which the visual button is encoded directly as a specific coding sequence composed of '0' and '1', related to two coding patterns. Here we employ a single-beam pattern with high gain and an scattering-reduction pattern for amplitude discrimination, respectively corresponding to '1' (high amplitude) and '0' (low amplitude). As a proof of prototype, we show the text wireless transmissions by mind from one operator to another within our EBCM communication system. The operator A, as the text transmitter, sends the letters by visually staring at the character button on the GUI of EBCM. When the target letter is decoded from the EEG signals, a coding sequence based on the ASCII codes is implemented on FPGA to switch the time-varying patterns.

In the encoding process, since the buttons representing the related text characters have the corresponding ASCII codes, the selected button is directly translated to the binary ASCII codes with the frame header "11111111110000", as illustrated in Fig. 2c. Then according to the final code, the metasurface reflects the high or low intensity to the space. In the decoding process, we firstly collect the spatial EM energy using the receiving channel, including a microstrip antenna embedded beside the metasurface, as shown in Fig. 2e, as well as a low-noise amplifier (LNA) and a high-speed analog-to-digital convertor (ADC) controlled by FPGA. The collected data stream is a series frame set, which represents the sampled intensity at the acquisition rate of 10 MHz. We use the decoding algorithm to locate the position of the frame header to determine the starting point of the data frame, as illustrated in Fig. 2f. Then the sampled data are transferred into the binary ASCII codes, and we display the text in the GUI (provided in Supplementary Note S2).

Four text sequences are sent and received successfully by mind using the EBCM platform, including "HELLO WORLD", "HI, SEU", "HI, SCUT", and "BCI METASURFACE". A recorded video is provided in Supplementary Movie S1. The average inputting time of each character is about 5s using the P300-based BCI by a skillful BCI operator. Since the programmable metasurface can achieve the "0/1" code transmitting speed of at least 1 Mbps, the maximum character transmitting speed for the metasurface is about $5\times10^4$ characters per second (20 bits each sequence). Hence the final text-transmitting speed is about 12 characters per minute. It is worth mentioning that the P300-based BCIs yield great accuracies and robustness among various noninvasive BCIs [5,6]. It is possible to improve the text input speed by applying some quick-spelling paradigms [7,37].

**Experiment implementation and results of wireless communications**

The detailed illustration of the communication system is depicted in Fig. 3a, in which the transmitting and receiving parts are marked in orange and green. In the transmitting part, the EEG signals are firstly detected and processed with the BCI devices and translated into the corresponding control signals of FPGA. The control signals follow the signal coding principle of the corresponding interface shown in Figs. 2b-2d. The FPGA executes the coding pattern arrangements and drives the PIN diodes to the desired states. In the receiving part, the microstrip antenna (MSA) beside the metasurface obtains the EM signals from the transmitter and sends it into the LNA and detector. The detector samples the analog amplitude, which is further converted to the digital codes for FPGA. The presented process is unidirectional but the communication system of EBCM is bidirectional since the transmitting and receiving fronts are respectively the metasurface and MSA, as illustrated in Fig. 3b.

The experimental scenario is exhibited in Fig. 3c, in which Operator A executes the text transmitting task and Operator B receives and reads the text. The distance between transmitting and receiving metasurfaces is about 1.3 m, in which the transmitting metasurface is excited by a broadband antenna with a distance of 0.3 m, and the receiver is an antenna integrated near the receiving metasurface, which is connected to an LNA and a high-speed detector, as well as an ADC, controlled by another FPGA. The received and demodulated letters and text are finally displayed on the designed GUI. The experiment process of the wireless text communications is recorded in Supplementary Movie S1.

Fig. 4a illustrates the processed EEG responses of the channel OZ to two stimulus types (target and nontarget) when the subject is spelling the word "HELLO". In each subfigure corresponding to the spelling of one character, for each of the two stimulus types, the event-related potential (ERP) waveforms are extracted by the time-locked average of the EEG signals across all flashes of the target or one of the nontarget buttons in a trial. Compared to the nontarget data, high amplitude is clearly observed at about 300 ms after the stimulus. According to the EEG signals, EBCM produces the amplitude-modulated EM signals of different letters using the ASCII code.

To fully demonstrate the wireless communications, we further provide 12 segments of the measured EM signals, including the letters 'H, E, L, O, B, C, I, S, M, E, T, A,' as showed in Figs. 4b-4d, where the high and low amplitudes respectively mean '1' and '0' in the ASCII code. The presented data are collected by an EM detector and then normalized. Each detection generates an amplitude pulse and numerous amplitude pulses to compose the presented data, where the high and low amplitudes respectively mean the codes 1 and 0. The ASCII codes of these letters are clearly observed according to Figs. 4b-4d.

**More functions of wavefront syntheses by EBCM**

To further demonstrate more functions of wavefront syntheses by the brain signals, we design three typical applications including the visual-beam scanning, multiple EM modulations, and coding pattern input. Here, we establish a demonstration prototype of EBCM, in which the metasurface is replaced by the light emitting diode (LED) version instead of PIN diode, as shown in Fig. 5a. Since the coding patterns on the metasurface directly determine the EM functions, we embed LEDs to intuitively visualize the pattern control in the EBCM verification system (see Supplementary Note S3). In the visual-beam-scanning scheme, we wish that the operator can direct the EM beam scanning accordingly using EBCM by simply staring at various directions in the sky, as depicted in Fig. 5b. We design a beam-scanning GUI in Fig. 2e, and the detailed descriptions are provided in Supplementary Note S4 with a sky background. We record a video of the visually-controlled beam scanning towards multiple directions, as presented in Supplementary Movie S2. We envision that this scheme can be further integrated with the augmented reality (AR) technique and finds more applications in adaptive mind-text wireless communications and intelligent radar detections.

To realize multiple EM modulations by EBCM, we design a specific interface to show that the operator can directly drive the EBCM for diverse EM functions (see Supplementary Note S4), including beam deflections, orbital-angular momentum (OAM) beam generations, and radar cross section (RCS) controls, as shown in Fig. 5c. The simulated results of these functions are presented in Figs. 5d-5m. For example, we illustrate the result of vertically-reflected single beam generated by the uniform phase pattern in Fig. 5d, and the results of three deflecting angles in Figs. 5e-g, where the simulated data clearly indicate the scattering directions of 15°, 30°, and 45°, suggesting great consistency with the measured directions marked in the red cross. More results are given in Supplementary Note S5. In the RCS level control, four RCS buttons from '01' to '04' will generate the scattered fields presented in Figs. 5i-l, showing the scattering levels of -15 dB, -12 dB, -9 dB, and -6 dB, respectively. For the OAM-beam generation, we exhibit two scattered fields of two OAM modes (+1 and +2) in Figs. 5h and 5m, in which the central amplitude null is clearly observed.

For experimental verification of the multiple EM modulations by EBCM, we recorded the controlling process of three representative EM functions, as shown in Supplementary Movie S3. The related coding patterns of 10 deflecting directions are given in Supplementary Note S6. The measured results presented in Supplementary Note S7 [36] show good agreements with the simulations in visual-beam scanning and multiple EM modulations. The related 30-channel EEG signals are also listed in Supplementary Note S8 [36], where the P300 waves are successfully captured from the EEG signals when the target button flashes. In addition, we perform the metasurface pattern coding scheme using EBCM. We show that the 2-bit phase coding patterns of the metasurface can be directly input from the brain signals to generate various EM functions. More details are illustrated in Supplementary Note S9 and Supplementary Movie S4.

**Discussion and conclusion**

We present an EBCM platform for direct mind-text wireless communications and diverse EM manipulations under non-invasive brain signals. We establish a new control manner from the operator consciousness to the metasurface pattern and realize EM functions by combining the P300 BCI device and the programmable metasurface. We propose and experimentally verify a wireless text-communication demo to directly transmit information from mind to mind based on EBCM between two operators. We show that the operator no longer needs any muscle-involved actions but only stares at the specific visual button for related sequential stimulus, which can be recognized by the EBCM and translated into the corresponding EM signals for communications. We also demonstrate three typical schemes with distinct functions, including visual beam scanning, multiple EM function switching, and metasurface pattern input, which contains more than 20 coding patterns for different single-beam scanning, multi-beam forming, OAM-beam generation, and RCS control. The presented work, combining the EM wave space and BCI, may further open up a new direction to explore the deep integration of metasurface, human brain intelligence, and artificial intelligence, so as to build up new generations of bio-intelligent metasurface systems.

## Methods and Materials

### The brain signal analysis based on P300 scanning

A 40-channel EEG amplifier NuAmps (Compumedics, Neuroscan, Inc., Australis) and a 30-channel EEG cap (LT 37) that followed the extended 10-20 system are used to collect the EEG signals, and the signals are referenced to the right mastoid. All electrode impedances are maintained below 5 kΩ during the data collection. Before manipulating the metasurface, the operator is instructed to perform a calibration stage, during which the EEG signals are collected to build up a training set, which is used to train a model for prediction in online manipulation stage. The calibration stage consists of 30 trials, and each trial consists of 10 rounds of button flashes. In each trial, the operator is instructed to focus on one target button, which is specified by the BCI program rather than freely determined by the operator.

The collected 30-channel EEG signals then undergo the preprocessing and feature extraction processes. Specifically, the EEG signals are firstly bandpass-filtered at 0.5-20 Hz. Then for each button flash (i.e., epoch) in a trial, a segment of EEG signal from 0 to 600 ms after the flash onset is extracted and baseline-corrected with a baseline extracted from the 200 ms before the flash onset. For each flash, the segment of EEG signal is subsequently down-sampled at a rate of 6. Since the sampling rate of the EEG signals is 250 Hz, there are 250Hz×600ms/6=25 sampling points for each channel per flash. Last, the segment of each flash is normalized along each channel using z-score normalization. Each obtained segment of signal, for example, segment corresponding to the flash of the $b$-th button in $r$-th round, forms a 30×25-dimensional matrix, and all rows of the matrix are concatenated together to form a vector, denoted as $\mathbf{x}_{r,b}$, as the feature vector of this fragment. To improve the signal-to-noise ratio (SNR), we average the feature vectors corresponding to the first $R$ rounds of each button in a trial as follows:

$$\overline{\mathbf{x}}_{R,b} = \frac{1}{R}\sum_{r=1}^{R}\mathbf{x}_{r,b}. \tag{1}$$

For each trial, a segment is labeled as a positive one if and only if the corresponding button is the target of the current trial, which indicates the presence of the P300 potential.

For each trial of the calibration data, we average the feature vectors over all the 10 rounds as in Eq. (1) to build up the training set, which is subsequently used to train a Bayesian linear regression model to decode the EEG segments, i.e., to distinguish the segments between with and without the presence of P300 potential.

In the online manipulation stage, to achieve a better balance between the accuracy and the speed of giving commands, the number of flash rounds is determined adaptively according to the confidence in making a decision. Similar to the calibration stage, the real-time collected EEG signals undergo the preprocessing and feature extraction processes to obtain feature vectors. After each round of button flashes, i.e., different values of $R$, the feature vectors are averaged as in Eq. (1). The averaged feature vectors corresponding to different commands are respectively fed into the trained decoder, and the outputs of the decoder are regarded as the confidence scores of the presence of P300 potential. When the score difference between the command with the highest score and the command with the second highest score is greater than a preset threshold (0.2 in this study), the command with the maximum score is sent; otherwise, the BCI proceeds to the next round of button flashes until the score difference reaches the threshold, or the number of flash rounds reaches the upper limit (10 rounds in this study).

**The coding and decoding methods on text communications**

In the text communication scheme, once the text from the operator's brain signal is detected by the BCI device, the FPGA should receive the text through the serial port and produce a binary sequence based on the ASCII code of the character. Taking character 'A' for example, the final coding sequence, composed of the frame header and the ASCII code of 'A', is encoded as '1111111000000001000001'. This coding sequence is also the switching sequence for FPGA, in which '0' and '1' respectively are the pattern for low and high reflection intensity. Here we employ an RCS reduction pattern and a single beam pattern as '0' and '1'. FPGA always monitors the serial port and switches the metasurface patterns for one sequence period when receiving a letter.

In the receiving part, the microstrip antenna integrated around metasurface undertakes the receiver to detect the transmitted energy. The coupled energy is firstly amplified by an LNA and sent into a detector with high accuracy. The detector performs sampling at 10 times the frequency of the modulator to recover the transmitted signals. The detected voltage is converted into digital within an ADC and then processed by a receiving FPGA. According to our coding rules, the decoded text is finally displayed on the screen of the operator B.

**EM experiment configurations**

The EM measurements are performed in a standard microwave chamber room, including the beam deflections, RCS reductions, and OAM-beam generations. The far-field measurement system consists of a rotatable table, two standard horn antennas, a signal generator (Keysight E8267D), a spectrum analyzer (Keysight E4447A), and an LNA. The feeding horn and the metasurface sample are both fixed on the rotatable table for far-field tests. In measuring the OAM phase distribution, the measurement is performed in a near-field measurement system, which is composed of a two-dimensional scanning frame, a vector network analyzer (Agilent N5230C), a probe and an LNA. In the communication scheme, a signal generating module (LMX 2594) and an LNA is applied to excite the metasurface. The receiving detector samples at the frequency of 10 MHz. It should be noted that the experimental verification of the EBCM control and the EM function modulation is not performed together in the chamber room since the BCI device and operator have interference to the EM measurement.

**Acknowledgements**

This work was supported in part from the National Key Research and Development Program of China (2017YFA0700201, 2017YFA0700202, and 2017YFA0700203), the Major Project of Natural Science Foundation of Jiangsu Province (BK20212002), the National Natural Science Foundation of China (61871127, 61735010, 61731010, 61890544, 61801117, 61722106, 61701107, 61701108, 61701246, 61631007, 61633010, 61876064, 62076099, 61731010, and 11874142), the State Key Laboratory of Millimeter Waves, Southeast University, China (K201924), the Fundamental Research Funds for the Central Universities (2242018R30001), the 111 Project (111-2-05), the Fund for International Cooperation and Exchange of National Natural Science Foundation of China (61761136007), the Key R&D Program of Guangdong Province (2018B030339001), the Key Realm R&D Program of Guangzhou (202007030007), the Guangdong Basic and Applied Basic Research Foundation (2019A1515011773), and the Pearl River S&T Nova Program of Guangzhou (201906010043). C.-W.Q. acknowledges the financial support from the grant R-261-518-004-720 from Advanced Research and Technology Innovation Centre (ARTIC).

**Data Availability:** The videos about the experiments of our EBCM are provided in the supplementary materials. The other data that support the plots within this paper and other findings of this study are available from the corresponding authors upon reasonable request.

**Author contributions**: T. J. C., C. W. Q., Y. L., Q. M., and W. G. conceived the research; Q. M., and W. G. designed the EBCM devices and relevant algorithms; Q. M., W. G., Q. X., L. D., T. G., Y. Z., X. G., T. Y., C. L., X. K., Z. G., and L. L. contributed to the experiments; Q. M., W. G., T. J. C., C. W. Q., and Y. L. prepared the manuscript; and T. J. C., C. W. Q., and Y. L. initiated and supervised the research. All authors contributed to the data analysis and writing of the manuscript, which was reviewed by all authors.

**Competing interests**: The authors declare no competing interests.

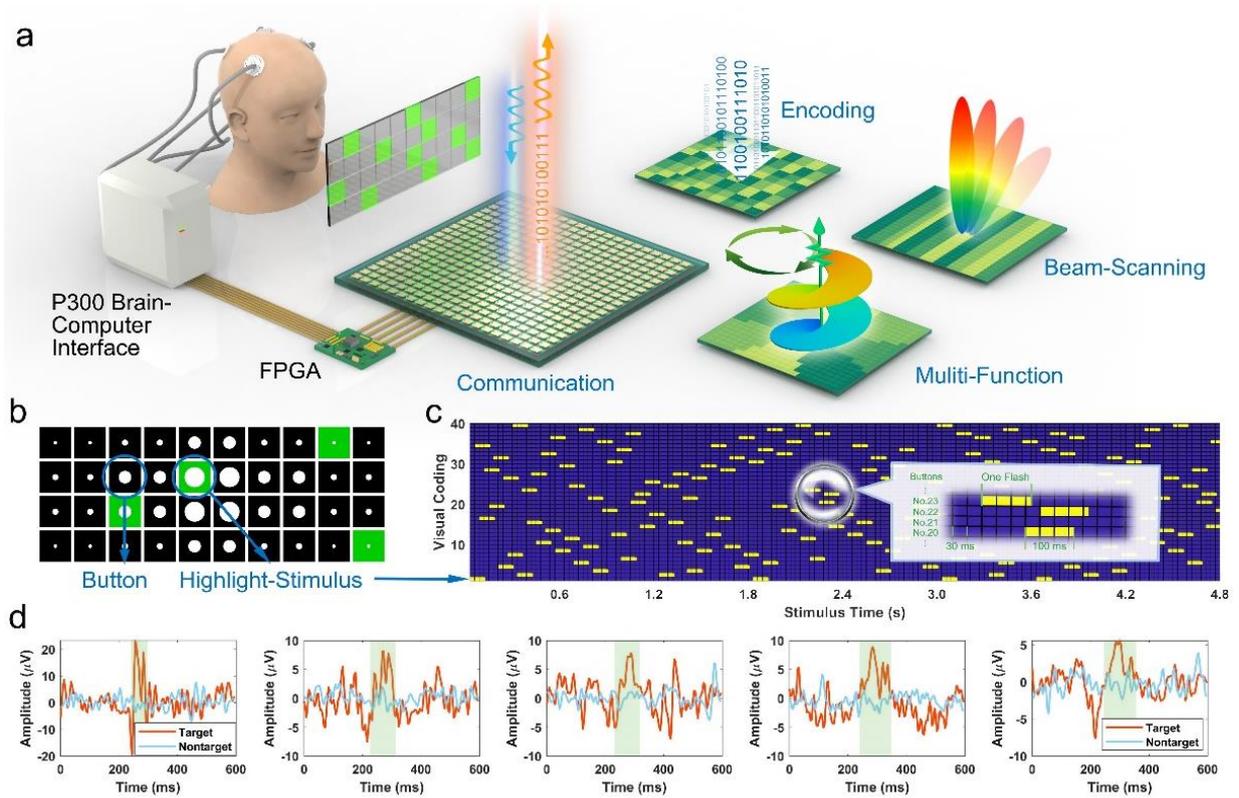

**Fig. 1. The EBCM platform.** (a) The system architecture of EBCM. The operator equipped with electrodes and P300 BCI device can directly instruct the metasurface with diverse EM functions under visual stimulus with the specific temporary coding sequences. Four typical schemes including brain-wireless communications, coding pattern encoding, beam scanning, and multi-function of EM modulations are demonstrated. (b) The graphical user interface of the beam deflection scheme. The buttons with different circles represent different beam scattering directions, where the highlight stimulus is green blocks. (c) Schematic diagram of the stimulus sequence, in which 40 rows represent the 40 buttons, and the yellow blocks mark the highlights of the buttons, each of which lasts for 100 ms. (d) The experimentally measured EEG signals of five button operations, where obvious amplitude peaks occur at about 300ms.

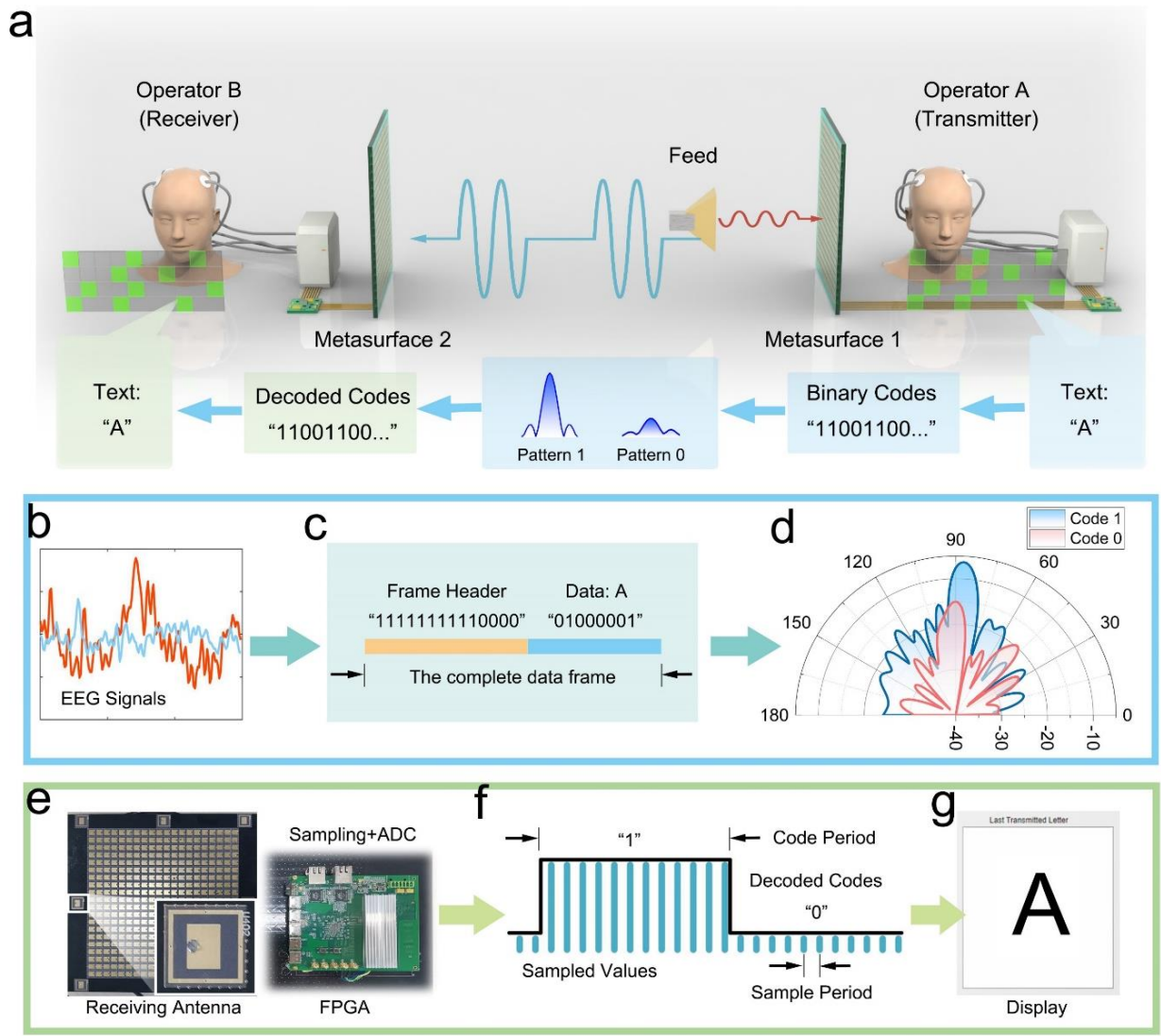

**Fig. 2. The wireless text-communication using EBCM.** (a) The system architecture of the text-communication system as well as the coding and decoding process. (b)-(d) The encoding process from EEG signals to the transmitted EM signals, where the EEG signals shown in (b) are first detected by BCI and translated into the digital sequence in (c) for wireless transmission, then radiated by the metasurface with different pattern amplitudes in (d). (e)-(g) The decoding process of wireless communication, where the antenna and FPGA in (e) firstly receive and sample the signals from space and convert them into the digital signals. The sampled data are discretized into 0/1 codes for decoding, as depicted in (f) and finally translated into the text for display.

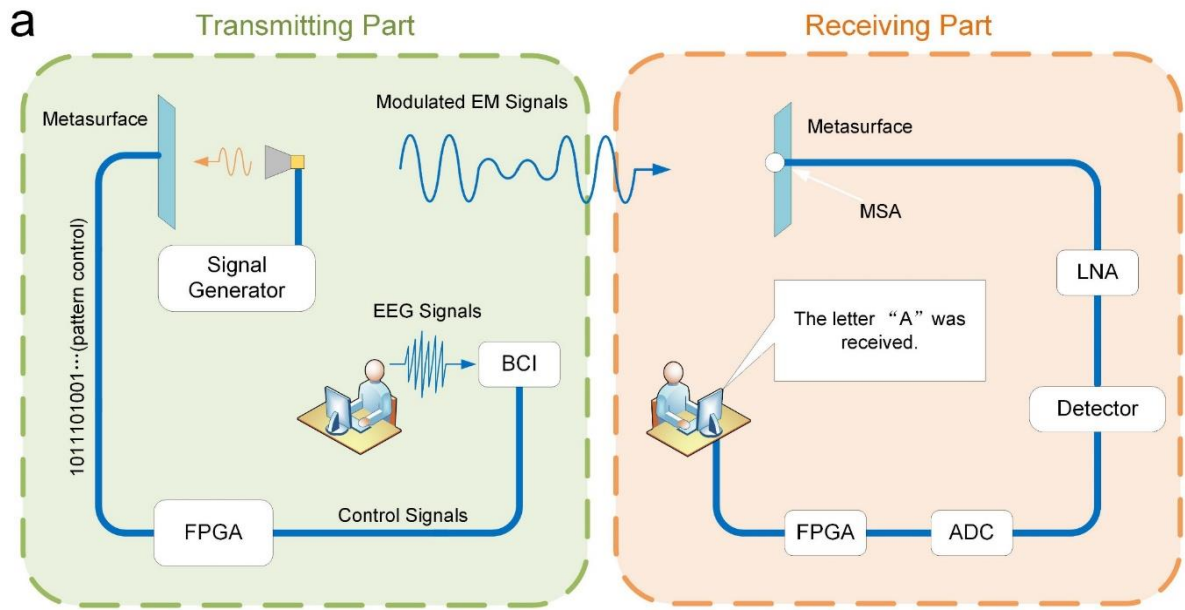

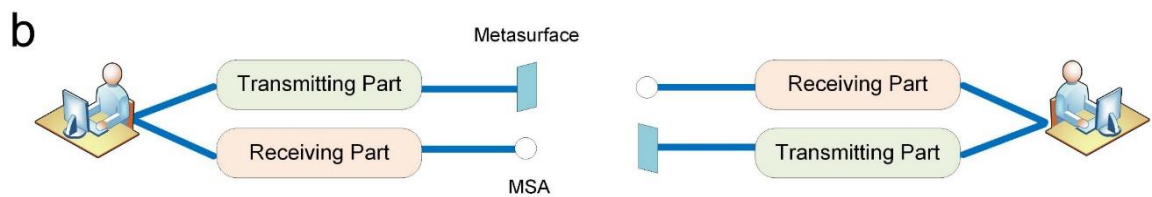

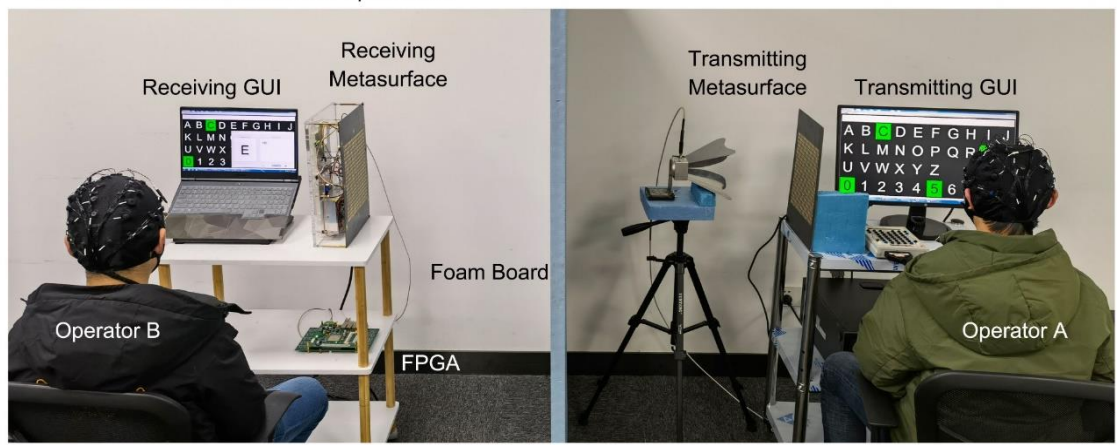

**Fig. 3. Experimental implementation of the wireless text-communication.** (a) The system architecture of the wireless text-communication experiment. (b) The illustration of the working mechanism of transmitting and receiving parts. (c) The experiment scenario of the wireless text-communications directly through EBCM, where a foam board is placed between two operators to verify the wireless properties.

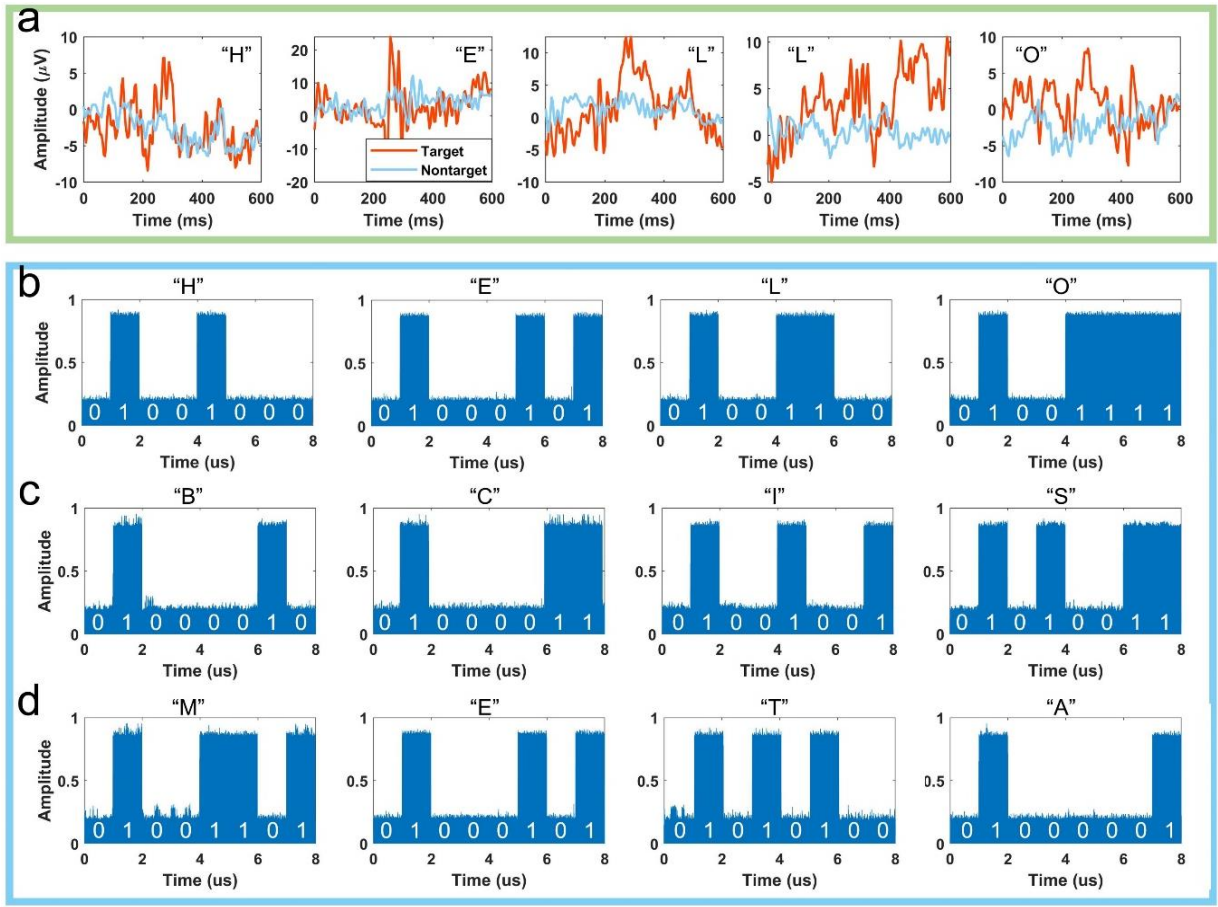

**Fig. 4. The experimental results of the wireless text-communications using EBCM.** (a) The experimentally measured EEG signals in the text communication scheme. The EEG segments corresponding to five letters "HELLO" are presented for demonstration. (b)-(d) The measured EM signals of the letters 'HELO', 'BCIS', and 'META'. The normalized amplitude-modulated signals present the ASCII codes of these letters, where high and low amplitudes mean '1' and '0'.

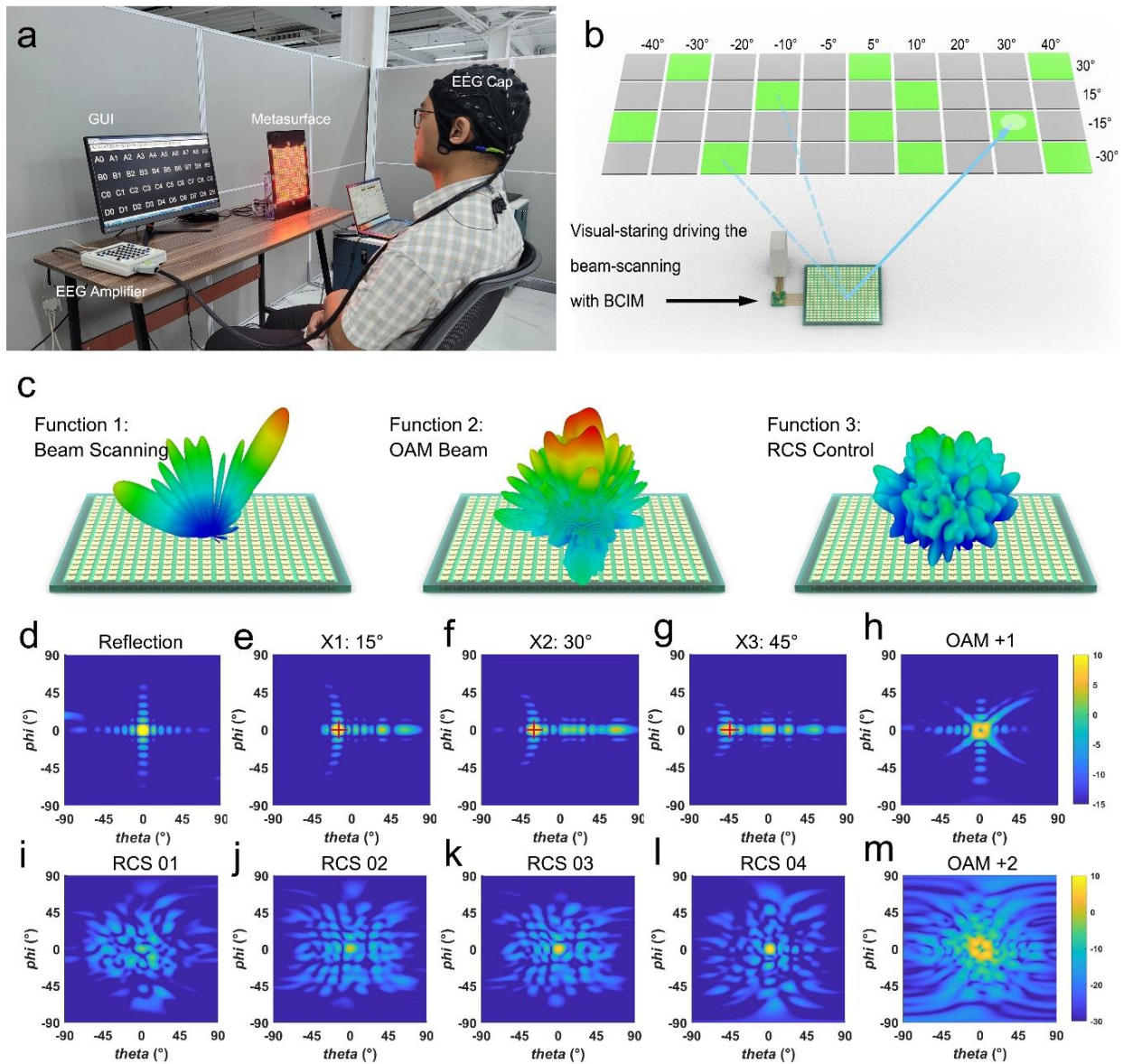

**Fig. 5. More functions of versatile wavefront syntheses using EBCM.** (a) The experiment photograph of the pattern encoding scheme. (b) The illustration of beam scanning by EBCM. The visual-staring of BCI operator directly drives the metasurface to adjust the scattering direction to the desired angle. Each button relates to a specific scattering direction as labelled. (c) Three typical EM functions including beam scanning, OAM-beam generation, and RCS control. (d)-(g) The simulated results of four kinds beam-scanning fields, where the measured result of the main direction is marked with the red cross, as listed in Supplementary Note S6[36]. The scattering beam is deflected from 0° to 45° along the x-axis. (g)-(j) The simulated field results of RCS control, where '01' to '04' indicate the four levels of reflection intensity. (k) and (l) The far-field simulation results of OAM modes +1 and +2. The central null is clearly observed.